# Optical Two-Centered Charges, Dipoles, and Conveyor-Belt Modes in Bipolar Coordinates.

J. STROHABER[1, *]

[1]*Department of Physics, Florida A&M University, Tallahassee, Florida 32307, USA*
**james.strohaber@famu.edu*

**Abstract:** We holographically generate optical modes associated with bipolar-coordinate solutions of the Helmholtz equation. Unlike conventional Gaussian beam families, these modes support multicenter phase discontinuities. We investigate and demonstrate that they support multicentered phase singularities, including optical dipoles and two-centered charge distributions. We further identify a family of conveyor-belt modes whose intensity structure follows an extended trajectory connecting the charge centers. Experimental generation using a spatial light modulator is presented and compared with theoretical predictions. These results establish bipolar-coordinate optical modes as a new class of structured light fields for singular optics, beam shaping, optical manipulation, and the controlled generation of multicentered topological structures.

## 1. Introduction

In 1960, Herwig Kogelnik and T. Li published a paper describing two fundamental families of free-space optical modes: the Hermite–Gaussian and Laguerre–Gaussian modes [1]. Since then, the properties and applications of these modes have been studied extensively, both experimentally and theoretically. More recently, a broader family of free-space modes was discovered and experimentally verified, encompassing both the Hermite–Gaussian and Laguerre–Gaussian mode families as special cases [2]. Among these beam families, the Laguerre–Gaussian modes have attracted considerable attention because they possess a well-defined amount of orbital angular momentum (OAM) [3,4]. This property has motivated extensive investigations into OAM-related nonlinear optical phenomena, including coherent transfer of OAM in high-harmonic generation [5], Raman sideband generation [6], second-harmonic generation [7], and optical spontaneous parametric down-conversion [8]. Other research on OAM-carrying beams have also found important practical applications in areas such as optical tweezers and spanners for particle manipulation [9,10], optical vortex coronography [11], Improved Imaging and Microscopy (STED) [12].

In this work, we holographically generate multicentered optical modes using computer-generated holograms displayed on a spatial light modulator. Our objective is to engineer structured light fields with multiple singularity centers for applications in optical trapping and particle manipulation [10]. Bipolar coordinates provide a natural framework for describing optical fields associated with two spatially separated singularities. Unlike conventional cylindrical and Cartesian coordinate systems, which center around a single origin, bipolar coordinates possess two geometric poles. This unique geometry enables the formation of structured optical fields exhibiting multicentered phase singularities, two-centered topological charge distributions, and optical dipole configurations. To our knowledge, experimental generation of structured optical modes based on bipolar-coordinate solutions of the Helmholtz equation has not previously been reported.

Section 2 develops the optical modes arising from solutions of the Helmholtz equation in bipolar coordinates and introduces the two-centered charge and optical dipole solutions. Section 3 presents a family of conveyor-belt modes obtained through a polar-like reformulation of the bipolar coordinates. Section 4 describes the experimental generation of these modes using computer-generated holograms and compares the measured intensity and phase distributions with theoretical predictions. Finally, Section 5 demonstrates that familiar structured-light families, such as the Hermite–Gaussian and Laguerre–Gaussian modes, emerge naturally within the bipolar-coordinate framework, providing a connection between conventional beam families and the multicentered modes introduced here.

## 2. Derivation of Bipolar Beam Modes

Because we wish to create two-centered charge distributions, a natural starting point is bipolar coordinates [13]. Bipolar coordinates are an orthogonal space with two points $(x, y) = (\pm a, 0)$ known as the poles where the coordinate system becomes singular. One set of circles of constant $\tau$ circle the poles and another set of circles of constant $\sigma$ go through the poles. The forward and inverse coordinate transformation between Cartesian $(x, y)$ and bipolar coordinates $(\tau, \sigma)$ are

$$x = a\frac{\sinh(\tau)}{\cosh(\tau) - \cos(\sigma)}, \qquad y = a\frac{\sin(\sigma)}{\cosh(\tau) - \cos(\sigma)} \tag{1a}$$

$$\tau = \frac{1}{2}\ln\left(\frac{(x+a)^2 + y^2}{(x-a)^2 + y^2}\right), \qquad \sigma = \arctan\left(\frac{2ay}{x^2 + y^2 - a^2}\right) \tag{1b}$$

The angular coordinate $\sigma$ is multivalued because the inverse bipolar transformation contains a branch cut. Restricting $\sigma$ to its principal branch, $-\pi < \sigma \le \pi$, places the branch cut along the line segment connecting the two poles, $y = 0$ and $-a < x < a$. Points approaching this segment from above and below correspond to $\sigma = \pi$ and $\sigma = -\pi$, respectively. Consequently, the entire line segment joining the poles is mapped onto the coordinate boundary $\sigma = \pm\pi$. In particular, the Cartesian origin $(x, y) = (0, 0)$ corresponds to $\tau = 0$ and lies on this coordinate boundary. This behavior is analogous to the polar angle in ordinary polar coordinates, where the negative *x*-axis corresponds to the coordinate boundary $\theta = \pm\pi$. In bipolar coordinates, the corresponding boundary is the finite line segment joining the two poles, represented by $\sigma = \pm\pi$. Figure 1 illustrates the coordinate curves of constant $\tau$ and $\sigma$. The solid curves correspond to $\tau = [-2, -1, -0.5,\ 0.5,\ 1,\ 2]$ and form circles surrounding the poles, whereas the dashed curves correspond to $\sigma = \left[-2.5, -2, -1.5, -1, -0.5,\ 0.5,\ 1,\ 1.5,\ 2,\ 2.5\right]$ and form circles passing through the poles. Positive values of $\tau$ generate circles centered in the positive *x*-half plane, whereas negative values of $\tau$ generate circles in the negative *x*-half plane. Similarly, positive values of $\sigma$ correspond to coordinate curves lying in the positive *y*-half plane and negative values in the negative *y*-half plane.

To better understand the results presented in the work, it is useful to examine the relationship between radial distances in Cartesian and bipolar coordinates systems. Throughout this paper $r^2 = x^2 + y^2$ will denote the usual radial coordinate in polar coordinates $(r, \theta)$, while $\rho^2 = \tau^2 + \sigma^2$ is used as a radial-like coordinate in bipolar variables $(\rho, \vartheta)$. The coordinate transformation in In Eq.1a reveals an inverse relation between Cartesian and bipolar radial distances. In the limit $(\rho, \vartheta) \to (0, 0)$, using first-order approximations for $\sinh(\tau)$ and $\sin(\sigma)$, and second-order approximations for $\cosh(\tau)$ and $\cos(\sigma)$, the Cartesian components become $x \approx 2a\tau / (\tau^2 + \sigma^2) \approx 2a\tau / \rho^2$ and $y \approx 2a\sigma / (\tau^2 + \sigma^2) \approx 2a\sigma / \rho^2$. Consequently, the radial distance is found to be $r \approx 2a / \rho$. This result demonstrates that the radial distances in the two coordinate systems are inversely related. As $\rho$ decreases, the corresponding physical distance $r$ increases. In other words, the neighborhood of $(\sigma,\tau)=(0,0)$ represents the far-field region in Cartesian space. As a consequence, features located at large Cartesian radii are compressed toward the origin of bipolar-coordinate space, whereas features near the Cartesian origin are mapped to larger values of $\rho$.

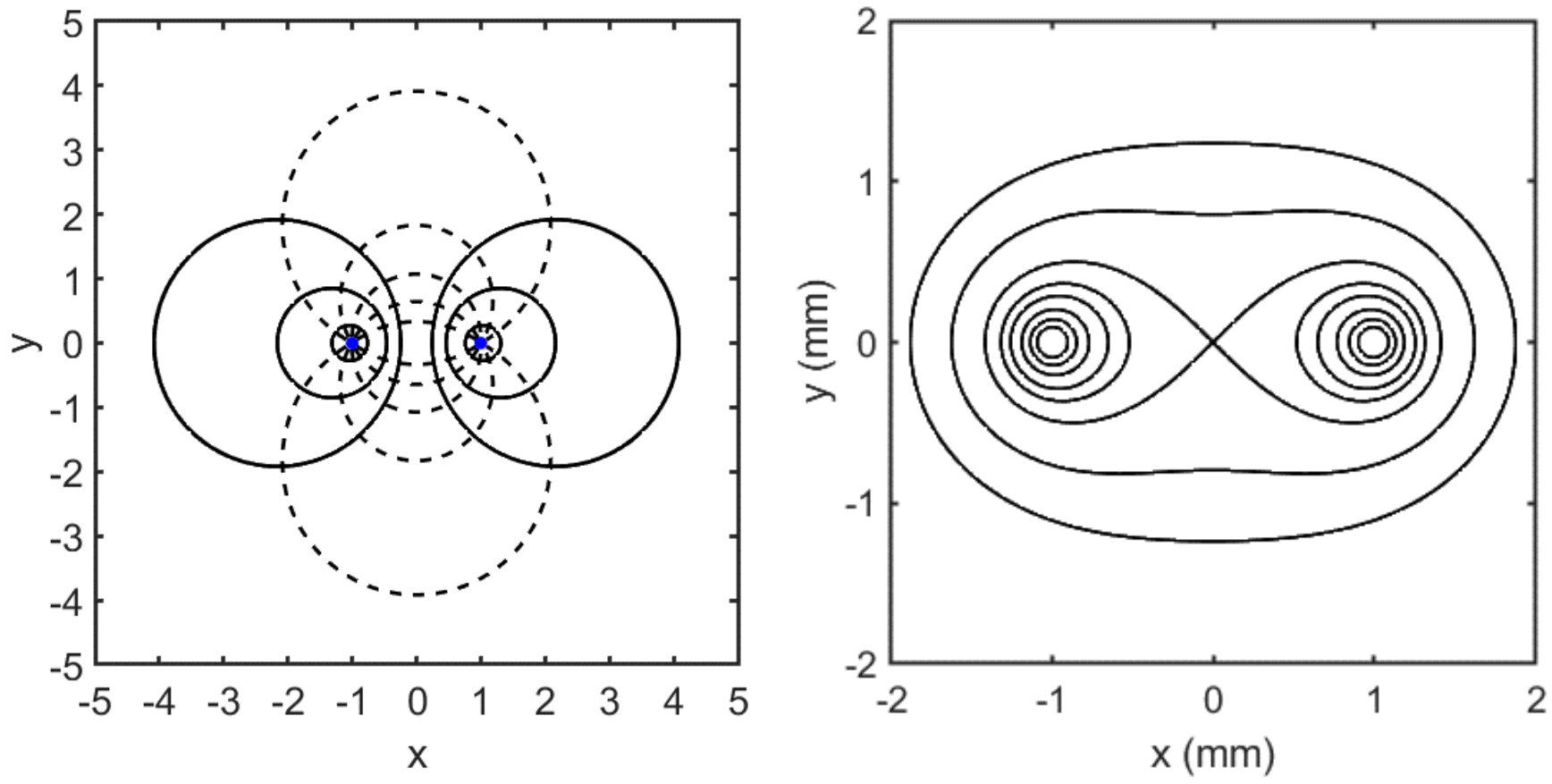

**Fig. 1.** (a) Bipolar coordinate system showing families of orthogonal coordinate curves. Solid curves represent constant $-\tau$ curves $\tau=[-2,-1,-0.5,0.5,1,2]$, which form circles enclosing the two poles. Dashed curves represent curves $\sigma=[-2.5,-2,-1.5,-1,-0.5,0.5,1,1.5,2,2.5]$, which form constant $-\sigma$ circles passing through both poles. The coordinate curves were calculated from the transformation given in Eq. (1a). (b) Contour plot of the refractive index for constant optical potential. The contour levels are [1 1.5 2 3 4 5 7 10 15 20].

To obtain optical modes in bipolar coordinates, we begin with Helmholtz equation $\nabla^2 E + k_0^2 n^2 E = 0$ in medium [14]. Expressed in bipolar coordinates,

$$\left[\frac{(\cosh(\tau)-\cos(\sigma))^2}{a^2}\left(\frac{\partial^2}{\partial\sigma^2}+\frac{\partial^2}{\partial\tau^2}\right)+\frac{\partial^2}{\partial z^2}+k_0^2 n^2(\tau,\sigma)\right]E=0\,, \tag{2}$$

where $k_0 = 2\pi/\lambda_0$ is the free-space wavenumber fixed by the laser wavelength, and $n(\sigma,\tau)$ is a graded refractive index profile to be determined. For propagating beams, we introduce the ansatz $E(\sigma,\tau,z)=\psi(\sigma,\tau)\exp(-i\beta z)$ which gives

$$\left[\frac{(\cosh(\tau)-\cos(\sigma))^2}{a^2}\left(\frac{\partial^2}{\partial\sigma^2}+\frac{\partial^2}{\partial\tau^2}\right)+k_0^2 n^2(\tau,\sigma)-\beta^2\right]\psi=0 \tag{3}$$

Here $\beta$ is the longitudinal propagation constant. Separation of variables can be achieved by engineering the refractive-index profile so that the bipolar scale factor multiplying the transverse Laplacian is canceled. Specifically, we require

$$k_0^2 n^2(\tau,\sigma)-\beta^2=\frac{(\cosh(\tau)-\cos(\sigma))^2}{a^2}V(\tau,\sigma)\,, \tag{4}$$

where $V(\tau,\sigma)=F(\sigma)+G(\tau)$ is the effective optical potential and $F(\sigma)$ and $G(\tau)$ are one-dimensional optical potentials arising from the refractive index distribution. Upon substitution of Eq. 4 into Eq. 3 results in the following separable equation

$$\left(\frac{\partial^2}{\partial\sigma^2}+\frac{\partial^2}{\partial\tau^2}\right)\psi+V(\tau,\sigma)\psi=0 \tag{5}$$

Defining $n_0=\beta/k_0$ as the uniform background refractive index, Eq (4) yields

$$n(\tau,\sigma)=\sqrt{n_0^2+\frac{(\cosh(\tau)-\cos(\sigma))^2}{k_0^2a^2}V(\tau,\sigma)} \tag{6}$$

The refractive-index profile therefore consists of a uniform background index $n_0$ together with a spatially varying contribution determined by the optical potential $V(\tau,\sigma)$. Various choices of the functions $F(\sigma)$ and $G(\tau)$ will determine the resulting family of guided modes. While no physical medium is known to possess this refractive-index profile, it was deliberately designed to achieve separability and generate optical modes that reflect the intrinsic geometry of bipolar coordinates.

The simplest case is obtained by choosing $F(\sigma)+G(\tau)=k^2$ where $k^2$ is a constant that should not to be confused with the optical wavenumber $k_0n$. Contours of the refractive index profile given by Eq. 6 for the parameters $k_0a=1$, $n_0=1$, and $F(\sigma)=G(\tau)=1$ are shown in Fig. 1. The refractive-index contours follow the natural geometry of the bipolar coordinate system. The two singular points of the coordinate transformation appear as distinct index towers, while the outer contours enclose both poles and form a figure-eight-like topology near the origin. This topology plays a central role in determining the structure of the supported optical modes. With this choice of the optical potential, the wave equation then reduces to

$$\left(\frac{\partial^2}{\partial\sigma^2}+\frac{\partial^2}{\partial\tau^2}\right)\psi+k^2\psi=0 \tag{7}$$

The resulting equation is mathematically equivalent to the time-independent Schrödinger equation for a particle confined in a two-dimensional box [15]. Consequently, separation of variables may be employed by using $\psi=T(\tau)S(\sigma)$ and introducing separation constants

$$\begin{aligned}\frac{1}{S(\sigma)}\frac{\partial^2}{\partial\sigma^2}S(\sigma)+k_\sigma^2=0\\ \frac{1}{T(\tau)}\frac{\partial^2}{\partial\tau^2}T(\tau)+k_\tau^2=0\end{aligned} \tag{8}$$

where $k^2=k_\tau^2+k_\sigma^2$. The corresponding solution is

$$\psi(\tau,\sigma)=\left[C\cos(k_\tau\tau)+D\sin(k_\tau\tau)\right]e^{ik_\sigma\sigma} \tag{9}$$

Since $\sigma$ is periodic over the interval $[-\pi,\pi]$, it naturally plays the role of an angular coordinate. For this reason, the boundary condition of the $\sigma$ coordinate follow from $S(\sigma+2\pi,\tau)=S(\sigma,\tau)$ which requires that $k_\sigma=n_\sigma$ where $n_\sigma=0,\pm1,\pm2,...$ and therefore $S(\sigma)=e^{in_\sigma\sigma}$. For the non-period coordinate $\tau$, we find that two physical boundary conditions are required for quantization and for a vanishing of the field at special infinity. We impose Dirichlet conditions on the curves $\tau=\pm\tau_0$ corresponding to circles surrounding the bipolar poles. Specifically, the boundary condition is the conditions $\psi(\tau = \pm\tau_0, \sigma) = 0$. Physically these boundaries may be interpreted as the walls of perfectly conducting wires. For the $\cos(k_\tau \tau_0) = 0$ term, this gives $k_\tau = (n_\tau + 1/2)\pi / \tau_0$ and for the $\sin(k_\tau \tau_0) = 0$ term we get $k_\tau = n_\tau \pi / \tau_0$ and Eq. 9 becomes

$$BP_{n_\rho n_\sigma} = \left\{ C\cos\left[\frac{(n_\tau + 1/2)\pi}{\tau_0}\tau\right] + D\sin\left(\frac{n_\tau \pi}{\tau_0}\tau\right)\right\} e^{in_\sigma \sigma} \tag{10}$$

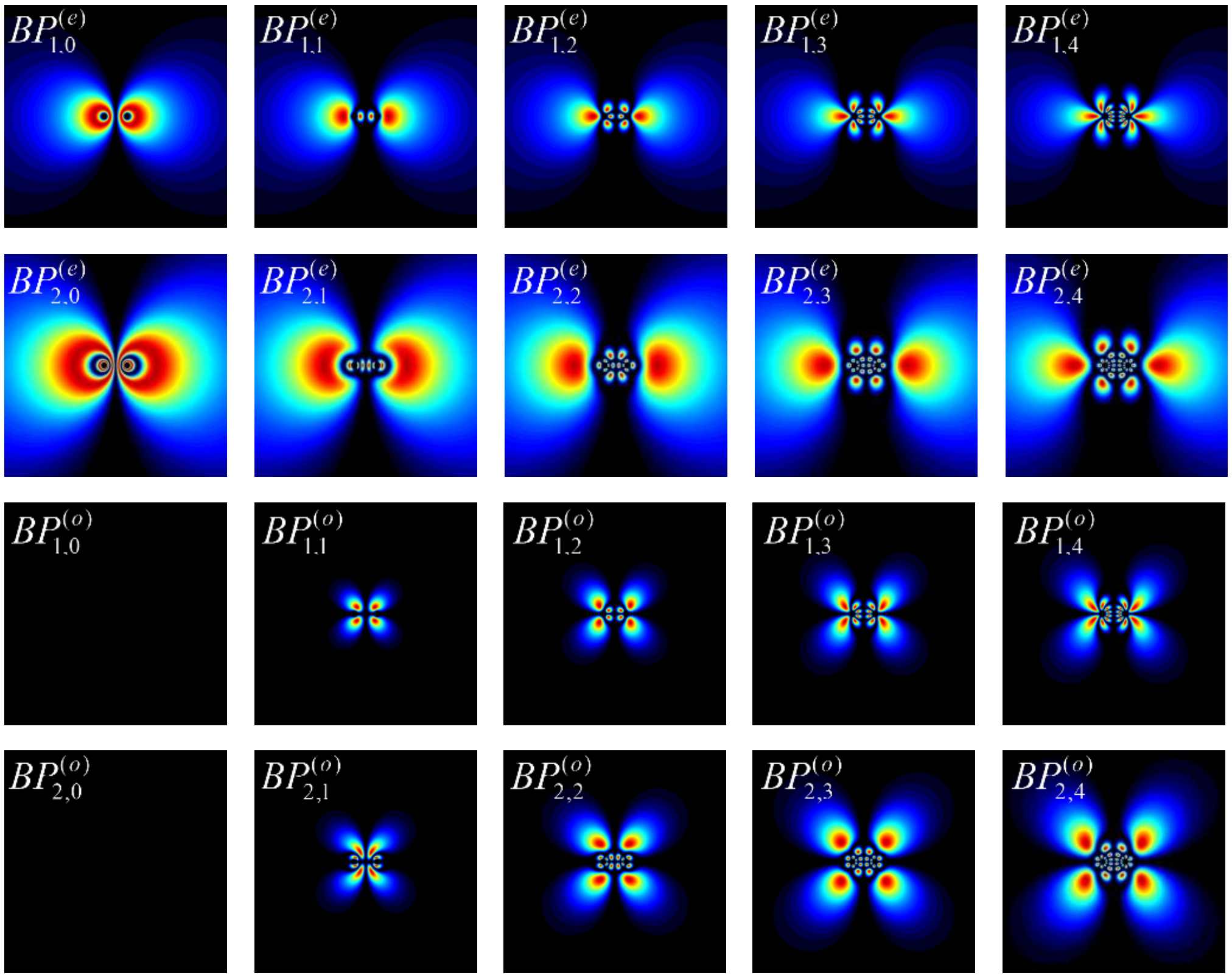

**Fig. 2. Calculated intensity distributions for even and odd bipolar modes.** Rows 1 and 2: even Bipolar modes with designated mode number $(n_\tau, n_\sigma)$. Rows 3 and 4: odd Bipolar modes with designated mode number $(n_\tau, n_\sigma)$. Odd modes have horizontal nodal line whereas the even modes do not.

Lastly, the asymptotic limit $r \to \infty$ corresponds to $\tau \to 0$ in bipolar coordinates. So requiring $\psi(r \to \infty) = 0$ means that in bipolar space $\psi(\sigma = 0, \tau = 0) = 0$ and the condition becomes $T(0) = 0$, which immediately eliminates the cosine family since $\cos(0) = 1$. Therefore, the sine family survives and our solution are the helical bipolar modes

$$BP_{n_\rho n_\sigma} = \frac{1}{\sqrt{\pi\tau_0}} \sin\left(\frac{n_\tau \pi}{\tau_0}\tau\right) e^{in_\sigma \sigma}, \tag{11}$$

where $n_\tau = 1, 2, 3, \ldots$ and $n_\sigma = 0, \pm 1, \pm 2, \ldots$. The requirement $\psi(r \to \infty) = 0$ eliminates half of the mathematically valid solutions; however, this condition is imposed to obtain modes that vanish at spatial infinity. The prefactor $1/\sqrt{\pi\tau_0}$ in Eq. 11 is a normalization factor over the domain $-\pi \leq \sigma \leq \pi$ and $0 \leq \tau \leq \tau_0$ in bipolar space. Two further solutions exist separating Eq. 11 into real and imaginary parts $BP^{(e)}_{n_\rho n_\sigma} = \mathrm{Re}(BP_{n_\rho n_\sigma})$ and $BP^{(o)}_{n_\rho n_\sigma} = \mathrm{Im}(BP_{n_\rho n_\sigma})$. These are the even and odd modes [16].

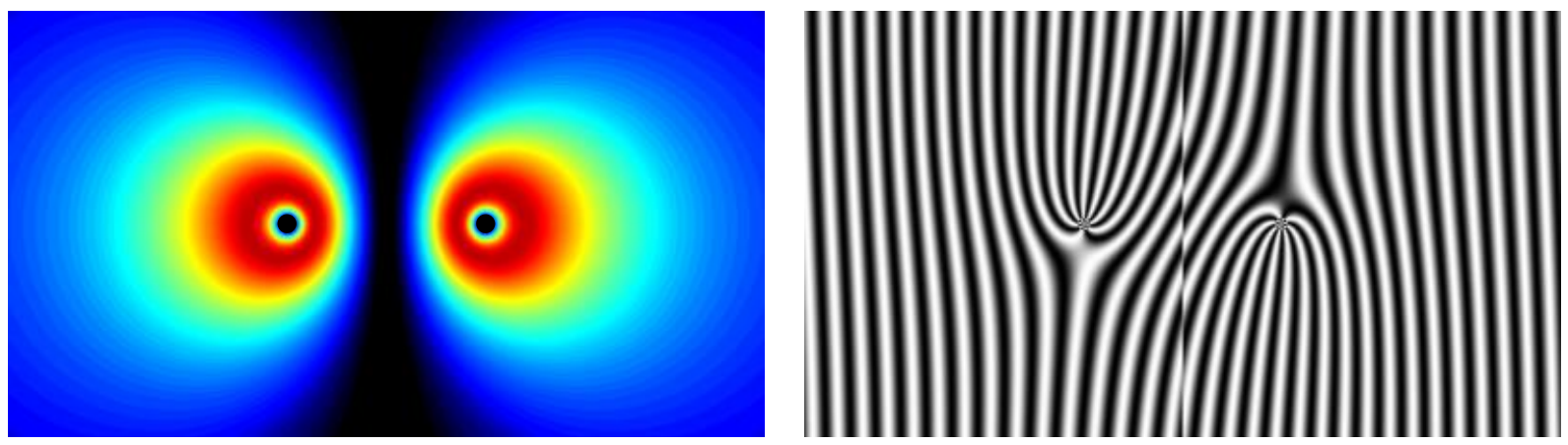

***Fig. 3. Theoretical optical dipole.*** *(a) Intensity distribution of the helical bipolar mode* $BP_{1,7}$*, exhibiting two spatially separated singularity centers that form an optical dipole. (b) Simulated interferogram obtained by interfering the field with a tilted plane reference wave. The opposite orientations of the fork dislocations indicate oppositely signed topological charges located at the bipolar poles, while the number of excess fringes determines the magnitude of each charge.*

Modal cross sections of the even $BP^{(e)}_{n_\rho n_\sigma}$ and odd $BP^{(o)}_{n_\rho n_\sigma}$ parts of Eq. 11 are shown in Fig. 2. Rows 1 and 2 are the even Bipolar modes with designated mode number $(n_\tau, n_\sigma)$ and Rows 3 and 4 are odd Bipolar modes with designated mode number $(n_\tau, n_\sigma)$. Odd modes have horizontal nodal line whereas the even modes do not. From the images, it is clear that the modal lobes circle around the poles. The Helical bipolar modes of Eq. 11 are constructed by the complex superposition of even and odd bipolar modes $BP_{n_\rho n_\sigma} = BP^{(e)}_{n_\rho n_\sigma} + iBP^{(o)}_{n_\rho n_\sigma}$. In particular, the mode $BP_{1,7} = BP^{(e)}_{1,7} + iBP^{(o)}_{1,7}$ along with its corresponding phase structure is shown in Fig. 3. The complex superposition produces an optical dipole consisting of two phase singularities of opposite topological charge located at the bipolar poles. This behavior is evident in the phase structure: around the left pole, there is an upright fork dislocation with seven extra fringes, whereas around the right pole the fork is inverted. The resulting field therefore exhibits a two-centered vortex structure with oppositely oriented local phase circulation.

## 3. Optical Conveyor-Belt

While the trigonometric bipolar modes exhibit multicentered singularity structures composed of spatially separated phase singularities of opposite topological charge, additional mode families emerge when the bipolar coordinates are recast into the polar-like variables $(\rho, \vartheta)$. In contrast to the dipole-like fields discussed above, the polar-like representation supports solutions in which the phase circulation has the same handedness about both poles, producing a continuous singularity structure that extends between the charge centers. This representation leads naturally to Bessel-type solutions that exhibit intensity transport along trajectories connecting the two poles. To obtain the optical conveyor belt modes, Eq. 10 may be transformed into polar-like representation by introducing the coordinates $\rho^2 = \tau^2 + \sigma^2$ and $\vartheta = \arctan(\sigma / \tau)$. Under this transformation, Eq. 6 reduces to the familiar Bessel equation yielding the solution

$$\psi = J_m(k\rho)e^{im\vartheta} \tag{12}$$

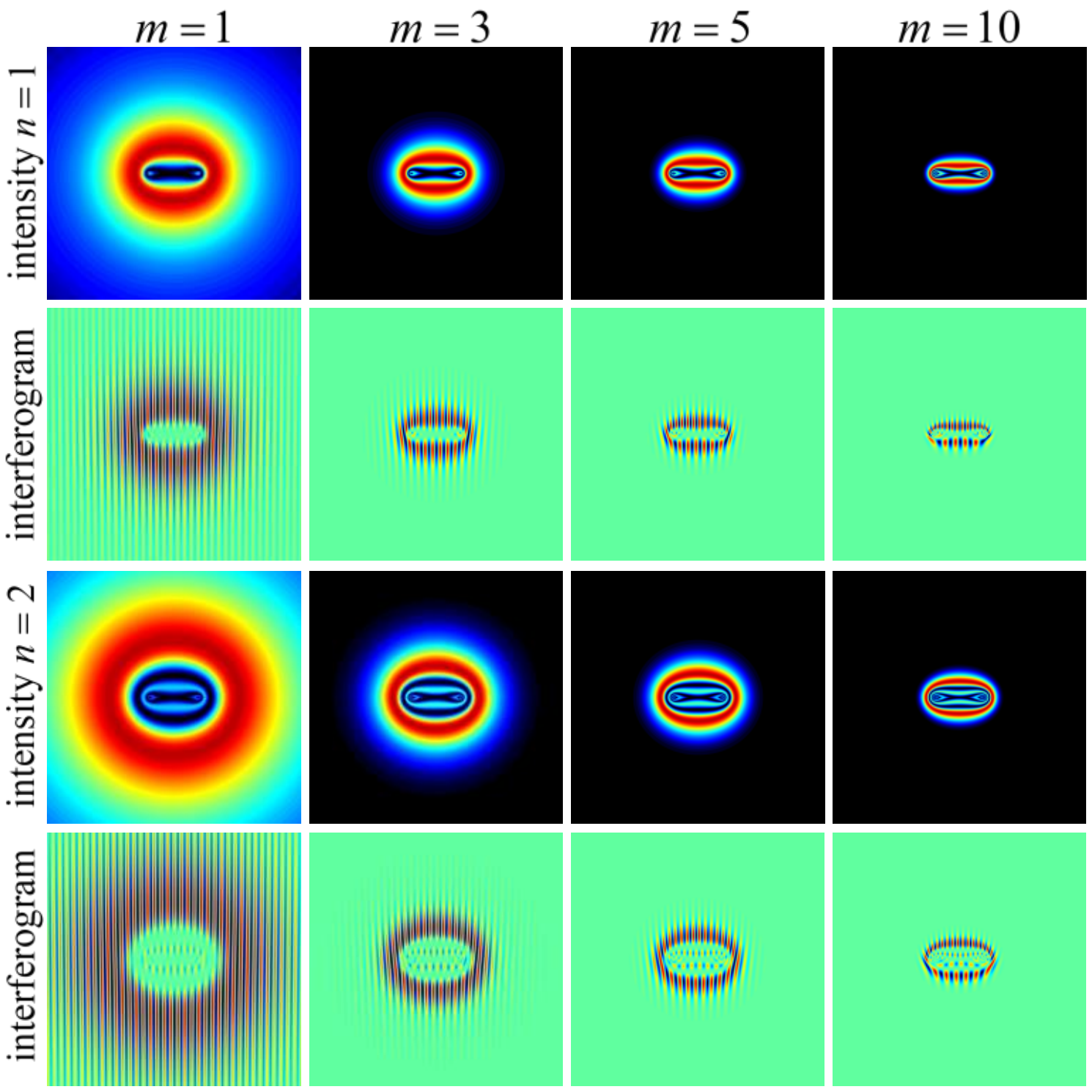

**Fig. 4. Conveyor-belt modes and corresponding interferograms:** The first and third rows show the calculated intensity distribution for radial mode numbers $n=1$ and $n=2$ respectively. The column correspond to increases values of *m*, which controls the amount of orbital angular moment in the beam. As *m* increases the beam becomes progressively more confined. The second and forth rows show simulated interferograms obtained using a tilted plane reference wave.

where $J_m$ is the Bessel function of the first kind of order $m$ [17]. As in the preceding section, the principal challenge lies in determining appropriate boundary condition. In conventional polar coordinates of a quantum particle in a circular box, the radial coordinate extends outward from the origin and the eigenvalues are obtained by imposing a boundary at a circular aperture of radius *R*, leading to the condition $J_m(kR)=0$. In the present coordinate system, however, the geometry is fundamentally different. The radial coordinate $\rho$ is defined in the $(\tau,\sigma)$ plane, and its contours are not physical circles in Cartesian space. Furthermore, the coordinate lines become increasingly compressed near the bipolar poles. Consequently, it is more natural to impose the boundary condition on an inner contour rather than on an outer boundary.

The quantized condition is found by imposing the Dirichlet condition $J_m(k\rho_0)=0$, and from this the allowed values of $k$ are found to be $k_{m,n}=j_{m,n}/\rho_0$ where $j_{nm}$ is the $n^{\text{th}}$ zero of $J_m(x)=0$. For example, under the condition $J_0(x)=0$ we have $j_{0,1}=2.4048$, $j_{0,2}=5.5201$, $j_{0,3}=8.6537$, and so forth.

At the Cartesian origin $(x,y)=(0,0)$, the inverse bipolar transformation Eq. 1b yields $\tau=0$ and $\sigma=\pm\pi$. Choosing the principal branch $\sigma=\pi$ and defining $\rho=(\tau^2+\sigma^2)$ gives $\rho_0 = \pi$ at the origin. Imposing the boundary condition $J_m(k\rho_0) = 0$ quantizes the transverse wavenumbers according to $k_{m,n} = j_{m,n} / \pi$ . The quantization condition results in $m = 0, \pm 1, \pm 2, \ldots$ and $n = 1, 2, 3, \ldots$. The solutions are therefore $\psi = J_m(j_{m,n}\rho / \pi)e^{im\vartheta}$. A second boundary condition is imposed at spatial infinity. Since $r \to \infty$ corresponds to $\rho \to 0$ we require that the field vanish as $\rho \to 0$. Because $J_m(0) = 0$ only for $m \neq 0$, the admissible modes satisfy $m = \pm 1, \pm 2, \ldots$ and $n = 1, 2, 3, \ldots$. The $m = 0$ mode is excluded because $J_0(0) = 1$ resulting a nonzero field amplitude at spatial infinity. The resulting conveyor-belt modes are shown in Fig. 4. In contrast to the localized charge and dipole solutions presented previously, these modes exhibit an extended intensity distribution that follows a continuous trajectory linking the two poles of the bipolar coordinate system. The field maxima form a chain of bright lobes resembling a conveyor belt, with the exact structure determined by the radial and angular mode numbers. The corresponding phase distributions reveal a coherent phase evolution along the trajectory. Their extended structure suggests potential applications in optical trapping and particle transport, where optical forces can be distributed along a prescribed path connecting multiple trapping sites.

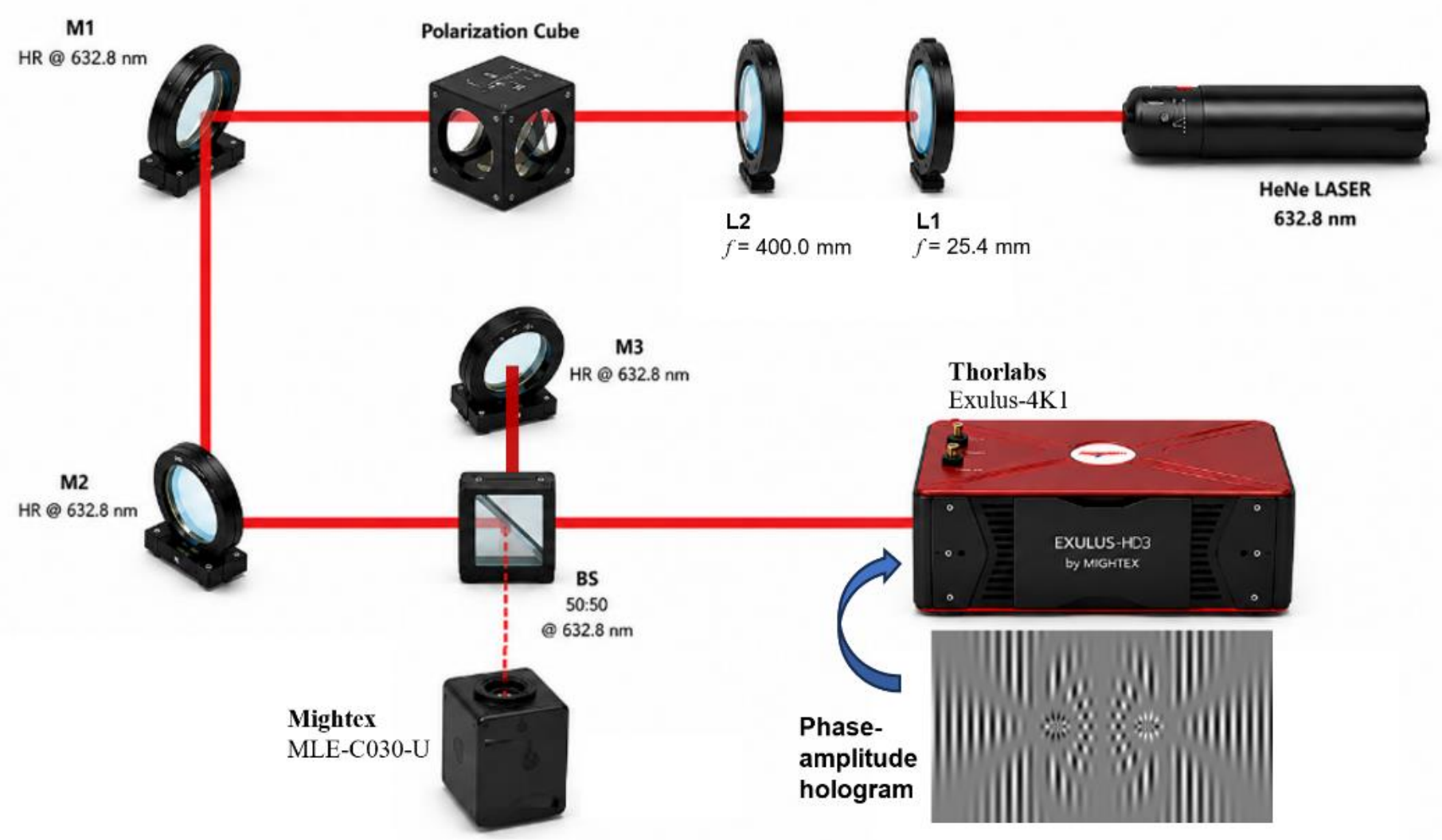

**Fig. 5. Experimental Setup.** A Michelson interferometer was used for characterization of holograms generated with the Exulus-4K1 spatial light modulator. Light from a polarization-stabilized HeNe laser is expanded with magnification of 15.75 and collimated before entering a polarization cube that filters unwanted residual vertical polarization. The beam is then sent through a 50:50 beam splitter to form the SLM and reference arms of the Michelson. The recombined light is detected with a collection optics and CCD camera.

## 4. Experimental Setup and Results

To experimentally investigate the calculated bipolar-coordinate modes, holograms corresponding to the even and odd mode indices shown in Fig. 2 were generated using the Exulus-4K1 spatial light modulator (SLM). Each complex field was encoded as an amplitude-modulated off-axis hologram [18], and the first diffraction order was isolated for characterization. The reconstructed intensity and interferometric phase distributions were recorded and compared with theoretical predictions.

The holograms were displayed on a Thorlabs Exulus-4K1 SLM with a resolution of 3840 by 2160 pixels and an operating wavelength range of 400–850 nm. Illumination was provided by a polarization-stabilized Melles Griot He–Ne laser (05-LPL-340) operating at 632.8 nm with an output power of approximately 3 mW and a beam diameter of 0.8 mm at the $1/e^2$ intensity level.

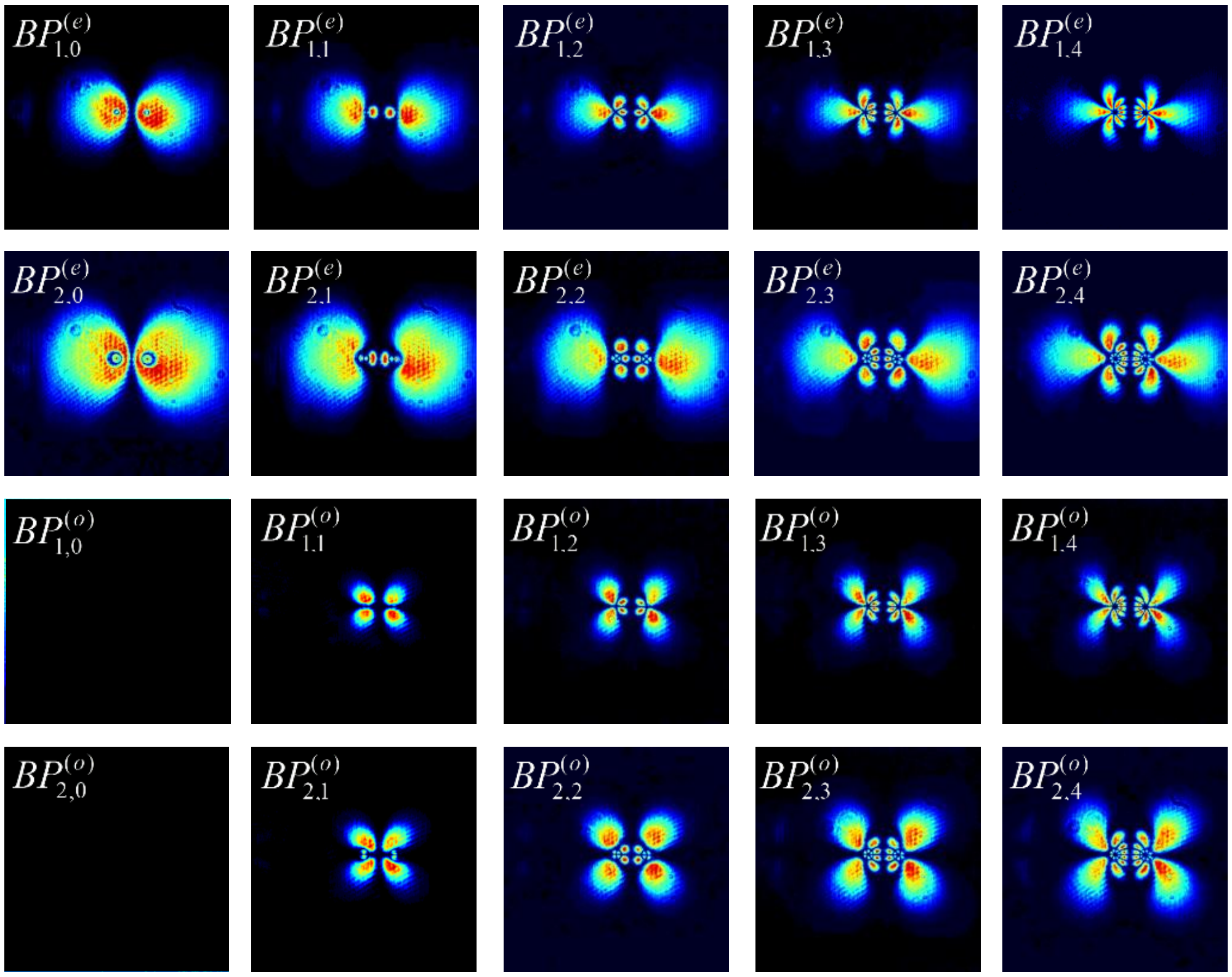

**Fig. 6. Experimental intensity profiels of even and odd bipolar modes.** The first two rows show the even modes $BP^{(e)}_{n_\tau,n_\sigma}$, while the last two rows show the odd modes $BP^{(o)}_{n_\tau,n_\sigma}$. The radial index $n_\tau$ increases from top to bottom and the angular index $n_\sigma$ increases from left to right. As $n_\sigma$ increases, additional intensity lobes develop in the region surrounding the poles, while increasing $n_\tau$ introduces additional radial structure. The modes were generated using phase-amplitude encoded computer-generated holograms displayed on a Thorlabs Exulus-4K1 spatial light modulator and recorded with a CCD camera. Each image has been cropped to have a dimension of 4.92 mm by 4.92 mm.

A Michelson interferometer, shown in Fig. 5, was used to characterize the generated holographic fields. The laser beam was expanded and collimated using a two-lens beam expander with focal lengths $f_1 = 25.4$ mm and $f_2 = 400.0$ mm, providing a magnification of $M = 15.75$ resulting in an output beam diameter of 12.6 mm. The beam then passed through a polarizing beam splitter cube to select the horizontal polarization required by the SLM. A 50:50 beam splitter divided the beam into the SLM and reference arms of the interferometer. Following recombination at the beam splitter, the resulting interference pattern was recorded using a Mightex MLE-C030-U CMOS camera equipped with an Aptina/Micron MT9T001 image sensor. The camera features a native resolution of 2048 × 1536 pixels with a 3.2 µm pixel pitch and a 6.554 mm × 4.915 mm active imaging area. All images were acquired in 1:2 decimation mode, resulting in an effective resolution of 1024 × 768 pixels.

In Fig. 6 even and odd bipolar modes are shown. Rows 1 and 2 are even modes with radial mode number $n_\tau = 1$ and $n_\tau = 2$ respectively, and angular mode numbers of $n_\sigma = 1$ to $n_\sigma = 4$

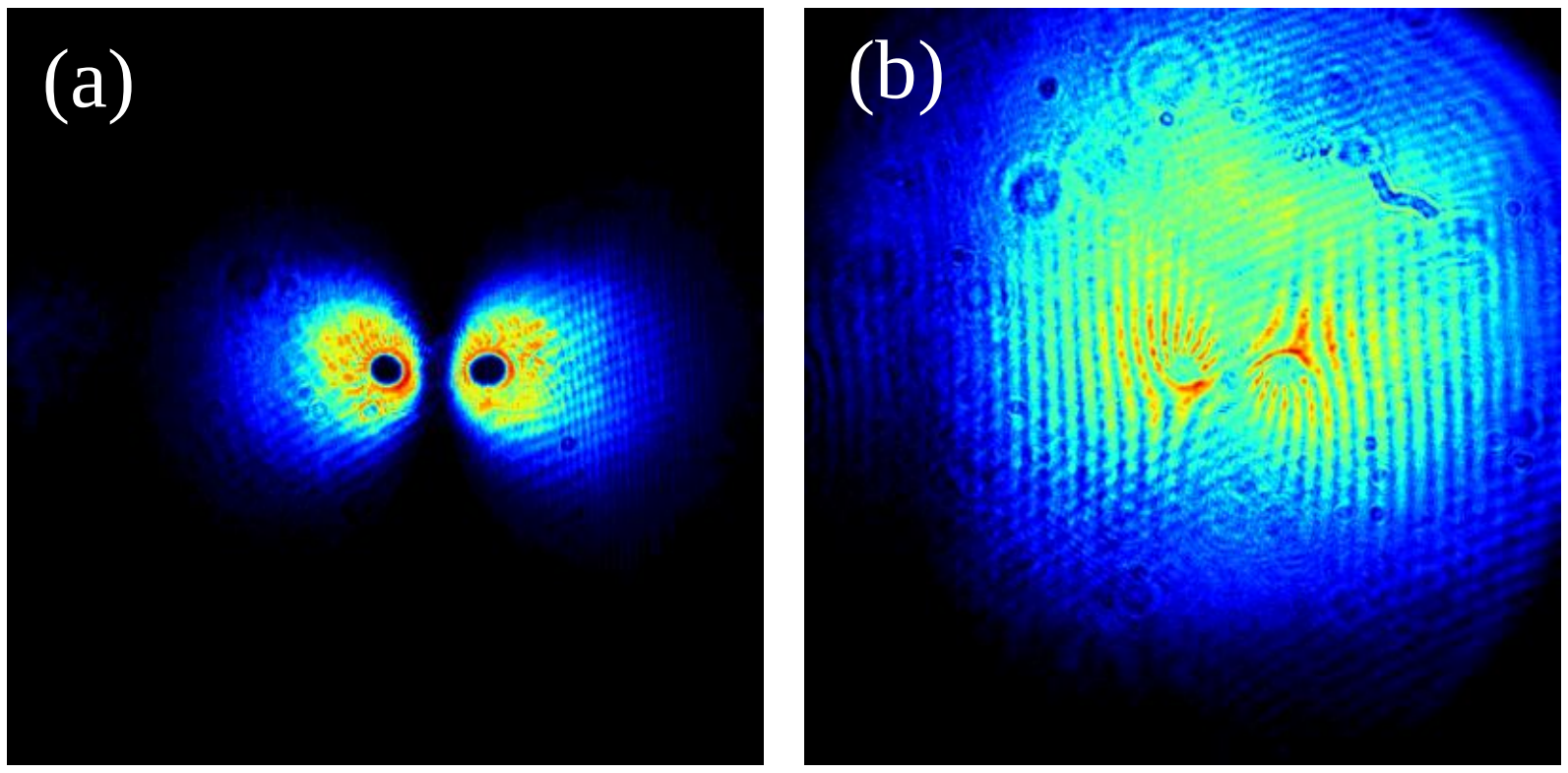

**Fig. 7. Experimental characterization of an optical dipole mode.** (a) Intensity distribution of a helical bipolar mode $BP_{1,7}$ . The field consists of two spatially separated singularity centers connected through the bipolar-coordinate structure. (b) Interferogram formed by interfering the field in panel (a) with a plane reference wave in a Michelson interferometer. The fork-like fringe dislocations exhibit opposite handedness, confirming the dipole nature of the mode and the presence of oppositely signed topological charges. The magnitude of each charge is determined from the number of excess interference fringes associated with the corresponding fork dislocation. The image size is 4.92 mm by 4.92 mm.

. It can be seen that as $n_\sigma$ increases the number of nodal lines following coordinate curves going through the poles increases. As the number $n_\tau$ increases the number of nodal circles surrounding the poles increases. The two poles are distinct and the modal lobes closely follow the geometry of the bipolar coordinate. The even and odd mode families exhibit distinct symmetry properties with respect to the horizontal axis passing through the poles. The even modes are symmetric under reflection about this axis, whereas the odd modes possess a nodal line along the axis and change sign across it. As a result, the odd modes display a four-lobed intensity distribution surrounding the poles, while the even modes concentrate a larger fraction of the optical power along the line connecting the two poles. The modal structure is strongly influenced by the geometry of the bipolar coordinate system. Unlike Hermite–Gaussian or Laguerre–Gaussian modes, whose nodal patterns are aligned with Cartesian or polar coordinate curves, the bipolar modes are organized around two singular points located at the coordinate poles. Consequently, the intensity maxima, nodal lines, and phase structure naturally conform to the two-centered geometry of the coordinate system. As the angular mode number $n_\sigma$ increases, additional nodal curves appear between the poles, producing progressively finer transverse structure. Increasing the radial mode number $n_\tau$ generates additional concentric nodal rings around each pole, causing the optical field to become more spatially localized. The combined action of $n_\sigma$ and $n_\tau$ therefore provides independent control over the angular and radial structure of the resulting optical mode.

Figure 7 presents the experimental generation and characterization of an optical dipole mode. The intensity distribution in Fig. 7(a) exhibits two spatially separated singularity centers located near the bipolar-coordinate poles. Unlike a conventional optical vortex, which contains a single phase singularity, the present field supports a pair of singularities embedded within the same optical mode. The intensity maxima surrounding each singularity follow the underlying geometry of the bipolar coordinate system, demonstrating the two-centered nature of the solution. The phase structure of the mode is revealed by the interferogram shown in Fig. 7(b). Interference with a plane reference wave produces two fork-like fringe dislocations with opposite orientation. The opposite handedness of the forks indicates that the singularities possess topological charges of equal magnitude and opposite sign. The field therefore constitutes an optical dipole, analogous to an electric dipole composed of positive and negative

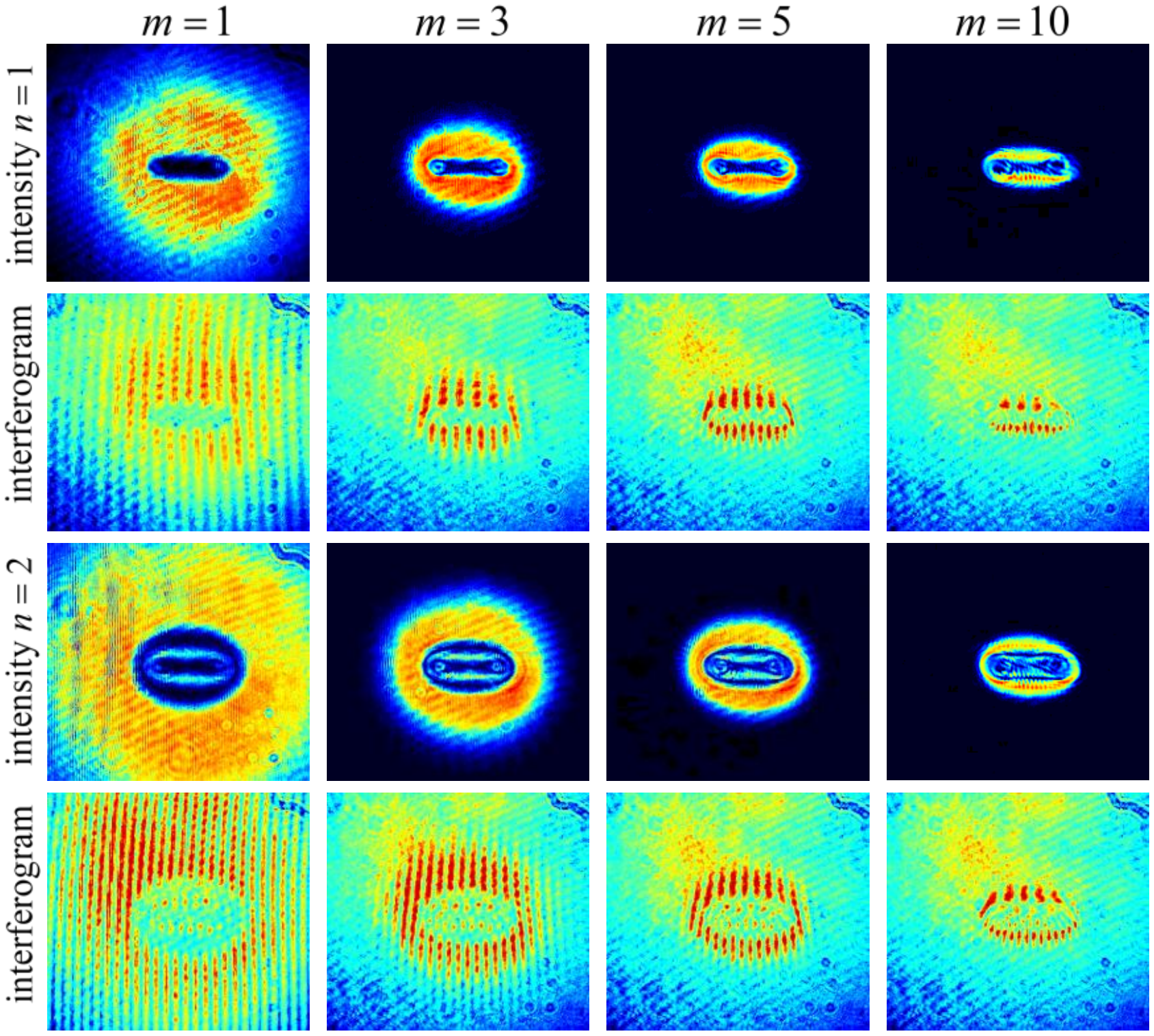

**Fig. 8. Experimental characterization of optical conveyor-belt modes.** Columns correspond to conveyor-belt orders $n = 1,3,5,10$. The first and third rows show the measured intensity distributions, while the second and fourth rows show the corresponding interferograms obtained with a plane reference wave in a Michelson interferometer. As the mode order increases, the optical field becomes progressively localized along the trajectory connecting the bipolar centers, forming an extended conveyor-belt-like structure. The interferograms reveal the associated phase distribution of the mode. The image size is 2.66 mm by 2.41 mm

charges separated by a finite distance. The bipolar coordinate system naturally provides a framework for describing such multicentered phase singularities.

The existence of oppositely signed topological charges leads to a phase structure that differs fundamentally from that of a conventional vortex beam carrying a single orbital-angular-momentum state. Instead, the optical field contains localized regions of opposite phase circulation centered on the two poles. These results demonstrate that bipolar-coordinate modes can support structured singularity distributions whose topology is determined by the geometry of the coordinate system rather than by a single on-axis vortex. The ability to generate and experimentally verify optical dipoles suggests that bipolar-coordinate modes provide a versatile platform for engineering multicentered singular optical fields and more complex topological charge distributions.

Figure 8 presents the experimental generation of optical conveyor-belt modes for several values of the angular mode number $m$ and for radial orders $n=1$ and $n=2$. Unlike the optical dipole modes discussed previously, which are characterized by localized phase singularities near the bipolar poles, these modes exhibit an extended intensity distribution that occupies the region connecting the two centers. The resulting optical field forms a continuous trajectory linking the poles, giving rise to the conveyor-belt-like appearance from which these modes derive their name. The measured intensity distributions reveal that increasing the angular mode number *m* progressively confines the optical field to the bipolar trajectory. For small values of *m*, the intensity occupies a relatively broad region surrounding the poles, whereas larger values of *m* produce a narrower and more strongly localized structure. This behavior is consistent with the increasing number of angular oscillations imposed by the bipolar-coordinate solution. The radial order *n* primarily controls the transverse structure of the field. The $n=1$ modes exhibit a single dominant intensity band connecting the poles, while the $n=2$ modes develop an additional radial node that divides the field into distinct intensity regions. Despite these differences, both mode families preserve the characteristic two-centered geometry of the bipolar coordinate system.

The corresponding interferograms provide information about the phase distribution associated with the conveyor-belt modes. Unlike the dipole mode of Fig. 7, which exhibit isolated fork dislocations corresponding to discrete topological charges, the conveyor-belt modes display an extended phase structure distributed along the trajectory connecting the poles. The interference fringes demonstrate that the optical field maintains a well-defined phase relationship throughout the entire conveyor-belt region, confirming that the observed intensity patterns arise from coherent bipolar-coordinate solutions rather than a simple superposition of independent localized modes.

These results demonstrate that bipolar-coordinate solutions support not only localized multicentered singularity states but also extended structured-light fields whose energy is concentrated along prescribed trajectories. Such modes may provide a useful platform for optical transport, particle guidance, and the controlled manipulation of matter along curved two-centered paths.

## 5. Hermite and Laguerre Gaussian Modes in Bipolar Coordinates

Having established the constant-potential modes, we now briefly show that other well-known structured-light families arise naturally within the same bipolar-coordinate framework. The optical potential shown in Eq. 5 can also take on the form $F(\sigma)=E_\sigma-\sigma^2$ and $G(\tau)=E_\tau-\tau^2$ which gives rise to

$$-\left(\frac{\partial^2}{\partial\sigma^2}+\frac{\partial^2}{\partial\tau^2}\right)\psi+\left(\sigma^2+\tau^2\right)\psi=E\psi\ , \tag{13}$$

which is Schrodinger's equation for the two-dimensional isotropic harmonic oscillator. Here the energy is $E=E_\sigma+E_\tau$. The solutions to Eq. 13 are well known in the optics community as the Hermite and Laguerre Gaussian modes.

$$\psi_{nm}(\sigma,\tau)=H_n(\sqrt{2}\sigma/\omega)H_n(\sqrt{2}\tau/\omega)e^{-(\sigma^2+\tau^2)/\omega^2} \tag{14a}$$

$$\psi_{p\ell}(\sigma,\tau)=(\sigma^2+\tau^2)^{|\ell|/2}L_p^{|\ell|}(\sigma^2+\tau^2)e^{-(\sigma^2+\tau^2)/2}e^{i\ell\tan^{-1}(\tau/\sigma)} \tag{14b}$$

Here represented in polar like coordinates in bipolar coordinate variables. The energies are $E=2(n+m+1)$ and $E=4\rho+2|\ell|+2$. The modal solutions in bipolar variables share a similar form to those that the constant optical potentials and is a topic of future research.

Figure 9 presents experimentally generated Hermite–Gaussian and Laguerre–Gaussian modes constructed in bipolar coordinates using computer-generated holograms displayed on the spatial light modulator. The measured intensity distributions exhibit the characteristic nodal structures associated with their corresponding mode orders in addition to reflecting the underlying two-centered geometry of the bipolar coordinate system. For the Hermite–Gaussian modes, increasing mode order leads to a greater number of localized intensity maxima concentrated near the coordinate origin and distributed symmetrically about the two poles. Likewise, the Laguerre–Gaussian modes display annular structures that arise naturally from the polar-like representation of bipolar coordinates. The experimentally observed patterns are in good qualitative agreement with the theoretical predictions and closely resemble the modal structures obtained from the trigonometric bipolar solutions of the previous section. These results demonstrate that conventional structured-light families can be reformulated within the bipolar-coordinate framework.

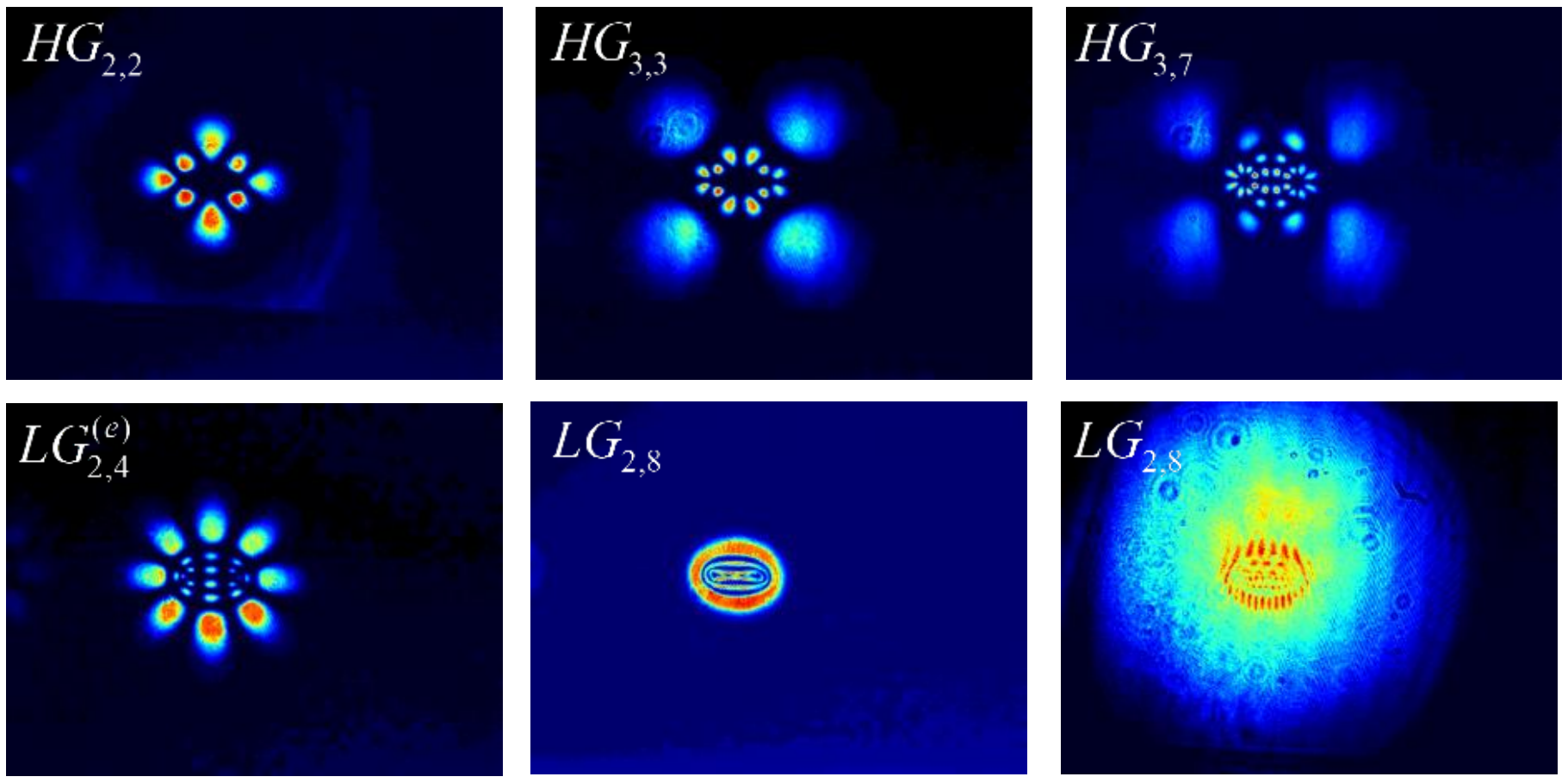

**Fig. 9.** Experimental intensity distributions of Hermite–Gaussian (HG) and Laguerre–Gaussian (LG) modes constructed in bipolar coordinates. The mode indices are indicated in the upper-left corner of each panel. The resulting fields exhibit multicentered intensity lobes and nodal structures that reflect the underlying bipolar geometry. Similar modal patterns are observed in the trigonometric bipolar modes of Fig. 6, demonstrating that conventional HG and LG beam families can be naturally reformulated within the bipolar-coordinate framework.

## 6. Conclusions

A brief discussion regarding the normalization of the bipolar modes is warranted. Although the modes are orthogonal and square-normalizable over the finite bipolar-coordinate domain used in the separation of variables, their behavior in Cartesian space is more subtle because the bipolar-coordinate Jacobian becomes singular at Cartesian infinity. As a result, not all bipolar-coordinate solutions remain square-normalizable when transformed to Cartesian coordinates. The even bipolar modes exhibit a logarithmic divergence in the Cartesian normalization integral, while the odd bipolar modes remain square-normalizable. The helical bipolar modes likewise exhibit a logarithmic divergence, whereas the conveyor-belt modes for $|m| \geq 2$ vanish sufficiently rapidly at Cartesian infinity to remain square-normalizable in Cartesian space. Modes whose field amplitude vanishes linearly at Cartesian infinity are not square-normalizable in Cartesian space, whereas modes that vanish quadratically or faster remain square-normalizable. In practice, however, all experimentally generated fields are produced over a finite aperture and therefore possess finite optical power. Consequently, these normalization considerations have little effect on the observed beam profiles and propagation dynamics reported here.

In this work, we have theoretically and experimentally demonstrated optical modes arising from solutions of the Helmholtz equation in bipolar coordinates. The resulting modes support multicentered singularity structures, optical dipoles consisting of oppositely charged phase singularities, and conveyor-belt modes whose intensity follows trajectories connecting the bipolar poles. Experimental generation using a spatial light modulator and Michelson interferometer showed good agreement with theoretical predictions. Bipolar-coordinate provides a natural geometry of optical fields containing multiple singularity centers. Unlike conventional Gaussian beam families, the geometry of the coordinate system itself determines the spatial arrangement of the optical singularities and nodal structures. The ability to generate two-centered charge distributions, optical dipoles, and conveyor-belt modes experimentally suggests that bipolar-coordinate optics may provide a useful platform for structured-light applications requiring controlled transport and manipulation along prescribed trajectories. Future work may explore propagation dynamics, optical trapping, and nonlinear interactions involving these multicentered fields.

**Disclosures.** The authors declare no conflicts of interest

**Data availability.** Data underlying the results presented in this paper are not publicly available at this time but may be obtained from the authors upon reasonable request.